# EMR Measurements of Field-Induced Superconductor $\lambda$-$(BETS)_2Fe_xGa_{1-x}Cl_4$

Y. Oshima [a*], E. Jobiliong [a], J. S. Brooks [a], S. A. Zvyagin [a], J. Krzystek [a], H. Tanaka [b], A. Kobayashi [c], H. Cui [d], H. Kobayashi [d]

[a] *National High Magnetic Field Laboratory, Florida State University, Tallahassee FL 32310, U.S.A.*

[b] *National Institute of Advanced Industrial Science and Technology, Tsukuba, Ibaraki 305-8568, Japan*

[c] *Research Centre for Spectrochemistry, Graduate School of Science, The University of Tokyo, Bukyo-ku, Tokyo 113-0033, Japan*

[d] *Institute for Molecular Science and CREST, JST, Myodaiji, Okazaki, 444-8585, Japan*

**Abstract**

We have performed high-field Electron Magnetic Resonance (EMR) measurements of $\lambda$-$(BETS)_2Fe_xGa_{1-x}Cl_4$ to clarify the role of $\pi$- and $d$-electrons in the paramagnetic (PM) and field-induced superconducting (FISC) phases. We found that this system has a non-negligible $\pi$-$d$ interaction that is very sensitive to the sample cooling rate. The $\pi$-$d$ interaction is weaker when the sample is rapidly cooled which might affect its FISC phase. The results of the organic alloy $\lambda$-$(BETS)_2Fe_xGa_{1-x}Cl_4$ ($x$=0.6) will be presented in this paper.

## 1. Introduction

$\lambda$-$(BETS)_2FeCl_4$, where BETS is bis (ethylenedithio) tetraselenafulvalene, is an interesting material that superconducts in a high magnetic field [1]. This field-induced superconductivity (FISC) can be explained by the Jaccarino-Peter compensation effect, where the internal magnetic field created by the $Fe^{3+}$ ($S$=5/2) moments through the exchange interaction is compensated by the external magnetic field [1-3]. Hence, the Zeeman effect, which normally destroys the superconductivity, is completely absent when the external field compensates the internal field [4]. Although the existence of strong $\pi$-$d$ interaction in this system is estimated from the mean field theory or the quantum oscillation measurement, no measurements that look directly at the electron spins have been performed in the high magnetic field region [5,6]. Therefore, we focused on the $\lambda$-$(BETS)_2Fe_xGa_{1-x}Cl_4$ salts because the FISC phase shifts to lower fields as the Fe contents $x$ decreases [3], and the electron magnetic resonance (EMR) measurement can be performed within our instrumental limitation. The results of the organic alloy $\lambda$-$(BETS)_2Fe_xGa_{1-x}Cl_4$ ($x$=0.6) will be presented in this paper.

## 2. Experimental

The crystal structure of the series of $\lambda$-$(BETS)_2Fe_xGa_{1-x}Cl_4$ alloys has triclinic symmetry [7]. The planar BETS molecules are stacked along the $a$-axis and have also intermolecular interactions along the $c$-axis, which form a 2D electronic structure. The insulating and magnetic anion layers are intercalated between these BETS layers. A finite exchange interaction between the $\pi$-electrons of BETS molecules and the $Fe^{3+}$ $3d$ electrons are expected due to the short inter-atomic distance between them [7]. The $Fe_xGa_{1-x}Cl_4$ ($x$=0.6) salt shows a metal-insulator transition at 6 K for zero magnetic field that is associated with the antiferromagnetic ordering of the $Fe^{3+}$ moments, and saturates above 8 T. The FISC phase appears above 10T when the magnetic field is applied along the $c^*$-axis as shown in Fig. 1. The samples are needle-shaped where the needle axis corresponds to the $c^*$-axis. High-field EMR measurements using the transmission method are performed with a 25T resistive magnet and a backward wave oscillator (BWO) as the light source. The dc magnetic field was carefully applied along the $c^*$-axis. The sample was cooled down from room temperature to liquid He temperature, with the cooling rate of 1 K/min. and 100K/min. for slow and rapid cooling, respectively. DPPH, which is a field maker, was used for the slow cooling data only.

---

[*] Corresponding author. Tel.: +1-850-644-6965; fax: +1-850-644-5038; e-mail: oshima@magnet.fsu.edu.

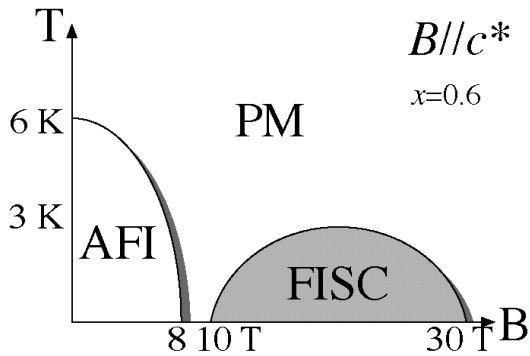

Fig. 1. The phase diagram of $\lambda$-(BETS)$_2$Fe$_x$Ga$_{1-x}$Cl$_4$ ($x$=0.6) for $B//c^*$-axis. PM, AFI and FISC denote for paramagnetic metal, antiferromagnetic insulator and field-induced superconductor, respectively.

## 3. Results and Discussion

Figure 2 shows the cooling rate dependence of typical EMR spectra for $\lambda$-(BETS)$_2$Fe$_x$Ga$_{1-x}$Cl$_4$ ($x$=0.6) at 4.2 K where the magnetic field is applied to the $c^*$-axis. Two typical spectra for 654.1 GHz and ~366 GHz are shown in Fig. 2. At both frequencies, it is clearly shown that two resonant absorption lines were observed for rapid cooling and just one for slow cooling. This suggests that the properties of the sample are different, depending on its cooling rate. This will be discussed later in this section.

The intensity and the linewidth of the two absorption lines for rapid cooling are different, and the resonance observed at the lower field is larger than the other. In principle, the absorption's intensity is proportional to the spin susceptibility. Hence, this difference is coming from the difference of the magnetic moments of the spins. The $g$-values are $g$~2.04 and 1.99 for resonance in lower and higher field, respectively. Considering these $g$-values and the absorptions' intensity, we can estimate that the resonance is coming from $d$-electron ($S$=5/2) and $\pi$-electron ($S$=1/2) for lower and higher resonance field, respectively.

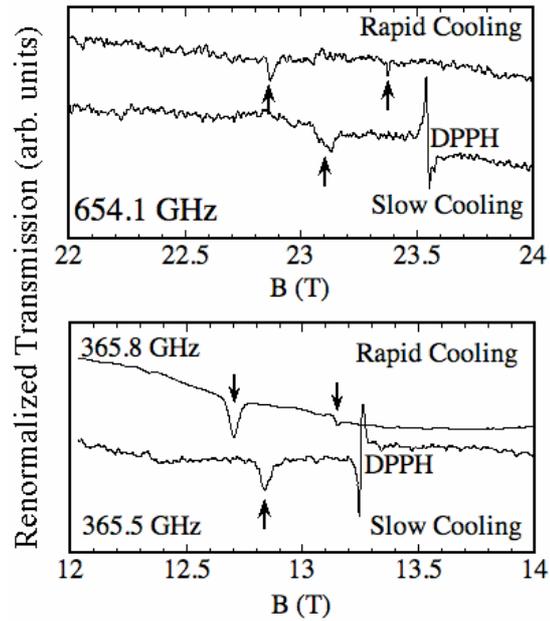

Fig. 2. The typical EMR spectra of $\lambda$-(BETS)$_2$Fe$_x$Ga$_{1-x}$Cl$_4$ ($x$=0.6) for 654.1 GHz (top) and ~366 GHz (bottom) at 4.2 K. The magnetic field is applied parallel to the $c^*$-axis. Each frequency shows its cooling rate dependence.

On the other hand, only a resonant absorption is observed for slow cooling. The frequency dependence of the effective $g$-value at around 2 K for slow cooling is shown in Fig. 3. Here, the effective $g$-values are obtained from the equation,

$$h\nu = g_{eff}\mu_B B \qquad (1),$$

where $\nu$ and $B$ is the frequency and the resonance field, respectively. It is clear that the effective $g$-value of this absorption is frequency dependent, and its $g$-value increases as the observing frequency is decreased. We think this is due to the magnetic anisotropy associated with the anions [8]. If we consider the spin Hamiltonian for this π-*d* compound, the resonance condition for EMR can be described as follows,

$$h\nu = g\mu_B B - 3D \quad (2),$$

where *g* is the intrinsic *g*-value of the system and *D* is the anisotropic spin term [8]. In this case, the frequency dependence of the effective *g*-value can be expressed as follows,

$$g_{eff} = g/(1 + 3D/h\nu) \quad (3).$$

It is clearly shown that the resonance plots in Fig. 3 are well fitted by eq.(3) and two parameters are obtained, $g \sim 2.01$ and $D \sim -3.3$ GHz $= -0.11$ cm$^{-1}$.

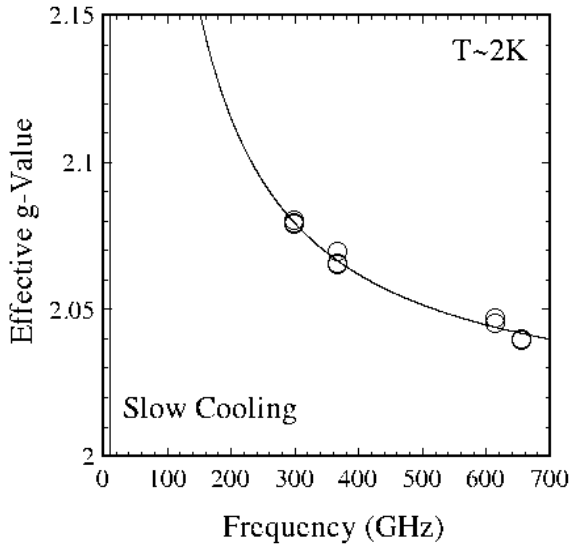

Fig. 3. The frequency dependence of the effective *g*-value at around 2 K for the slow cooling. The solid line shows the theoretical curve fit with $g \sim 2.01$ and $D \sim -3.3$ GHz.

The temperature dependence of the spectra for slow cooling at 613.8 GHz is shown in Fig. 4. The single absorption shifts to the higher field as the temperature increases. Although the resonance shift is generally due to the spin-orbit interaction, the existence of the interaction between the conduction electron and the localized spin can also causes the *g*-shift in the case of magnetic metals [9]. Thus, we believe that non-negligible π-*d* interaction exists in the system. More detailed discussion will be published elsewhere [8].

It is also worthy to note that there is no significant *g*-shift or linewidth broadening of the resonance when crossing the FISC phase boundary (i.e. at ~3 K). We think that the sample is only partially superconducting at this temperature and the resonance is coming from the paramagnetic domains.

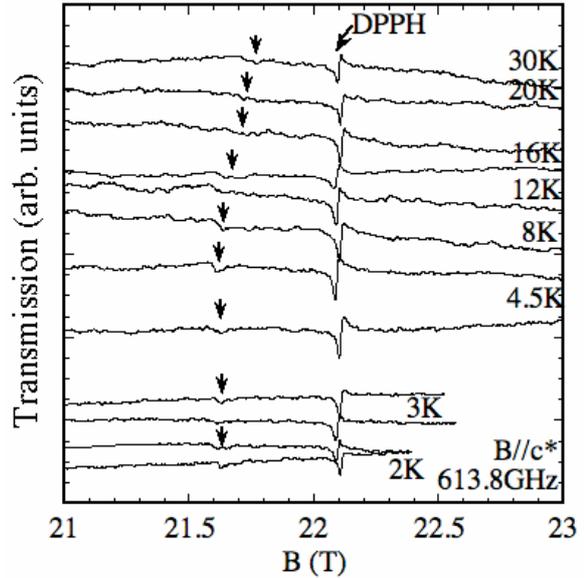

Fig. 4. The temperature dependence of the typical spectra for 613.8 GHz where the magnetic field is applied along the *c**-axis. The resonant absorption is represented by the arrow and the strong signal at ~22.1 T is a field marker (i.e. DPPH).

Finally, we briefly mention about the cooling rate dependence of the sample. In general, if there exists a strong exchange interaction between the π- and *d*-electron, each resonance should merge into one resonance. These resonances can be split when the Zeeman energy exceeds the exchange interaction, $2J \sim \Delta g \mu_B H$ where $\Delta g$ is the difference of the *g*-value between the two spins [10]. The exchange interaction *J* is estimated to be ~15 K from the mean field theory [5]. Then the Zeeman energy is around 625 GHz which is much higher than the observed frequency. Therefore, it seems quite normal that only a resonant absorption is observed for slow cooling. However, two absorption lines were observed for rapid cooling which suggests that the exchange interaction is smaller than for the slow cooling. The minimum frequency, that the two absorption lines were observed, was 306.4 GHz which suggests that the exchange interaction is less than ~7 K. Unfortunately, we could not obtain the exact exchange interaction due to the instrument restriction. The cooling rate dependence of the exchange interaction might come from the quenching of the donor molecules and the anions. It is possible that the inter-atomic distance between BETS molecule and $Fe^{3+}$ could change depending on the cooling rate. A careful X-Ray study is needed to verify that.

## 4. Conclusions

We have performed high-field EMR measurements on the organic $\pi$-$d$ electron system $\lambda$-$(BETS)_2Fe_xGa_{1-x}Cl_4$ ($x$=0.6). The observed signals for slow cooling have shown behavior that is related to a significant $\pi$-$d$ interaction. However, this exchange interaction between $\pi$- and $d$-electron is weakened when the sample is rapidly cooled. This suggests that the external field might not compensate the internal field and destroys the superconductivity. That may be the reason why the FISC phase does not appear for rapid cooling [11]. Finally, we did not observe any significant change in the EMR signal at high fields between the normal and FISC ground states. This may be a result of the nature of the FISC state, which, in the Jaccarino-Peter scenario, would allow flux to penetrate to the bulk sample.

## Acknowledgement

The author Y.O. acknowledges Dr. S. Uji and Prof. H. Ohta for helpful discussions. This work has been funded by NHMFL/IHRP 5042 and NSF-DMR-0203532. The NHMFL is supported by a contractual agreement between NSF and the state of Florida.

## References

[1] S. Uji, H. Shinagawa, T. Terashima, T. Yakabe, Y. Terai, M. Tokumoto, A. Kobayashi, H. Tanaka and H. Kobayashi, Nature 410 (2001) 908.
[2] L. Balicas, J.S. Brooks, K. Storr, S. Uji, M. Tokumoto, H. Tanaka, H. Kobayashi, A. Kobayashi, V. Barzykin and L. Gorkov, Phys. Rev. Lett. 87 (2001) 067002.
[3] S. Uji, T. Terashima, C. Terakura, T. Yakabe, Y. Terai, S. Yasuzuka, Y. Imanaka, M. Tokumoto, A. Kobayashi, F. Sakai, H. Tanaka, H. Kobayashi, L. Balicas and J.S. Brooks, J. Phys. Soc. Jpn. 72 (2003) 369.
[4] V. Jaccarino and M. Peter, Phys. Rev. Lett. 9 (1962) 290.
[5] T. Mori and M. Katsuhara, J. Phys. Soc. Jpn. 71 (2002) 826.
[6] S. Uji, C. Terakura, T. Terashima, T. Yakabe, Y. Terai, M. Tokumoto, A. Kobayashi, F. Sakai, H. Tanaka and H. Kobayashi, Phys. Rev. B 65 (2002) 113101.
[7] H. Kobayashi, H. Tomita, T. Naito, A. Kobayashi, F. Sakai, T. Watanabe and P. Cassoux, J. Am. Chem. Soc. 118 (1996) 368.
[8] Y. Oshima, E. Jobiliong, T. Tokumoto, J.S. Brooks, S.A. Zvyagin, J. Krzystek, H. Tanaka, A. Kobayashi, H. Cui and H. Kobayashi, cond-mat/0404645.
[9] R.H. Taylor, Adv. Phys. 24 (1975) 681.
[10] P.W. Anderson, J. Phys. Soc. Jpn. 9 (1954) 316.
[11] S. Uji, private communication.